\documentclass{sig-alternate-05-2015}
\usepackage{subfigure}
\usepackage{multirow}

\begin{document}






%

\title {The Study of Transient Faults Propagation\\
in Multithread Applications}

%
%
%
%
%

\numberofauthors{2} 
%

\author{
%
%
\alignauthor
Navid Khoshavi\titlenote{Navid Khoshavi is a Ph. D. student in department of electronic engineering and computer science at University of Central Florida. He received his M.S. from Amirkabir University of Technology (AUT) (Tehran Polytechnic), Iran, in 2012.}\\
       \affaddr{University of Central Florida}\\
       \email{nkhoshavi@eecs.ucf.edu}
\alignauthor
Armin Samiei\titlenote{Armin Samiei is a M. S. student in department of Computer Science at University of Central Florida. He received his B.S. from University of Shiraz, Iran, in 2012.}\\
       \affaddr{University of Central Florida}\\
       \email{samiei2@gmail.com}
}

\maketitle
\begin{abstract}
Whereas contemporary Error Correcting Codes (ECC) designs occupy a significant fraction of total die area in chip-multiprocessors (CMPs), approaches to deal with the vulnerability increase of CMP architecture against Single Event Upsets (SEUs) and Multi-Bit Upsets (MBUs) are sought.  In this paper, we focus on reliability assessment of multithreaded applications running on CMPs to propose an adaptive application-relevant architecture design to accommodate the impact of both SEUs and MBUs in the entire CMP architecture.  

This work concentrates on leveraging the intrinsic soft-error-immunity feature of Spin-Transfer Torque RAM (STT-RAM) as an alternative for SRAM-based storage and operation components. We target a specific portion of working set for reallocation to improve the reliability level of the CMP architecture design. A selected portion of instructions in multithreaded program which experience high rate of referencing with the lowest memory modification are ideal candidate to be stored and executed in STT-RAM based components. We argue about why we cannot use STT-RAM for the global storage and operation counterparts and describe the obtained resiliency compared to the baseline setup. In addition, a detail study of the impact of SEUs and MBUs on multithreaded programs will be presented in the Appendix.

\end{abstract}


\section{Introduction}
In order to keep performance improvement rates within a given power budget, the ITRS technology roadmap recommends the movement toward employing many-core processors in devices offering reduced power consumption and execution time. To maximize the beneficial of using many-core processors, thread level parallelism has been introduced as an inevitable counterpart in multicore programming \cite{Gizopoulos2011}. Meanwhile, the advances in CMOS technology have provided reduction in transistor size and voltage levels which results in significant increase of transient fault occurrence in the microprocessors. In particular, given roughly 50\% of chip is
occupied by memory structure, the existing memory module becomes highly susceptible to soft errors  \cite{ref:parma_2011}. 

Soft errors also referred as Single Event Upsets (SEUs) induced by energetic particles penetrate the silicon substrate and generate electron-hole pairs along their tracks. If the generated electrons collected into cell junction is larger than critical charge ($Q_{C}$), it can flip the cell state \cite{khoshavi2011control}, \cite{khoshavi2015facta}. The smaller device size and power supply reductions have severely increased the impact of SEU on deep-submicron technology, as they reduced aggressively the critical charge of memory cells \cite{khoshavi2012two}, \cite{khoshavi2011_SIES}. Thus, the memory cells have become more sensitive to atmospheric neutrons as well as to alpha particles which are created by unstable isotopes in the materials of a chip. 

Furthermore, the technology scaling and high precision manufacturing techniques have also decreased the separation distance between two adjacent memory cells which results in a single particle strike passing through adjacent cells in a row flip more than one cell. This phenomenon is called Multiple-bit Upset (MBU) \cite{gill_2005}. The main issue regarding the MBU handling is that the existing Error Correcting Codes (ECC) are not able to handle MBUs due to unpredictable behavior of the impact of soft errors when they flip more than one memory cell. To maintain acceptable reliability levels, the state-of-the-art MBU protection technigues have recently received significant attention to protect modern multicore architectures from the potential write and read failures introduced by soft errors. 

One intuitive solution is to replace traditional vulnerable SRAM-based memory technology with soft-error-immune memory modules such as Spin-Transfer Torque Magnetic Random Access
Memory (STT-RAM). The STT-RAM offers high density, low standby power, nonvolatility and soft error resiliency \cite{ref:xie_2013}\cite{ref:li_2015}. Recent research shows that the intrinsic immunity of STT-RAM to soft errors cause this device gets influenced by several order lower soft errors compared to SRAM \cite{tehrani_2010}\cite{sun_2012}. 

The authors of \cite{sun_2012} proposed an architectural radiation-induced soft error resilient solution for L1 cache through using STT-RAM as alternative for traditional SRAM. However, this work only concentrates on the impact of soft errors in L1 cache while the lifetime of a L1 cacheline is extremely short which means the error caused by particle strikes may not have enough time to be either consumed by CPU  or  propagated to the lower level of memory hierarchy. On the other hand, the authors of \cite{ref:xie_2013} studied the impact of using 3D stacked STT-RAM caches on the reliability of the whole cache hierarchy. They proposed a set of configurations for cache hierarchy and compared the results in respect of performance, power consumption and reliability. The obtained results show that the replacing memories with a STT-RAM alternative can significantly mitigate soft errors while offering slight performance improvement. 

In this work, we show that not every memory component in the processor core is required to be replaced by STT-RAM because the following reasons:
\begin{itemize}
\item The frequency of instruction call is not similar for every single instruction of the program. For example, a 95\% of a program execution time may be spent on an iteration which implies that this portion of program is more susceptible to soft errors if the generated error be consumed by CPU before masking the error through a write operation. 
\item The instructions running on CPU show various sensitivity to soft errors. For example, if a bit cell in register file of a sensitive instruction flipped due to particle strike, the error is immediately propagated into the multithreaded program which may result in the rapid crash of the program. 
\item The STT-RAM suffers from long write latency and high write energy which impose extra overhead to run write-intensive workloads in terms of performance reduction and high dynamic energy consumption.    
\end{itemize}

The ideal instructions to benefit from soft-error-immune STT-RAM memory component are those instructions which experience high rate of referencing with the lowest memory modification. Thus, a comprehensive study is required to recognize \textit{highly sensitive instruction to soft errors either to SEU or to MBU}, and \textit{frequently called instructions with the lowest memory modification} in the system.
To protect aforementioned instruction typeset, we propose to use STT-RAM memory module to maintain the reliability levels of the system in an acceptable error margin.
Accordingly, the trade-off among reliability, performance, and power consumption of the proposed technique compared to traditional methodologies will be explored.   



The outline of the rest of the paper is as follows: we examine the runtime behavior of multithreaded programs in Section 2. The proposed execution model assumed in this paper will be presented in Section 3. The experimental results will be discussed in Section 4. Finally, we conclude the paper in Section 5.

\section{The Runtime Program Behavior}
The runtime program behavior is required to be investigated to help us to determine which functions or instructions are ideal candidates to be mapped on soft error resilient STT-RAM components.

\subsection{The Frequency Call of Instructions}
In order to investigate the instruction references in multithreaded programs, we used Visual Studio 2013 Profiler Instrumentation with no special optimizations on Intel i7 with 8 GBs of RAM. The data values and ranges in the report are named "Elapsed Inclusive Time \%", "Elapsed Exclusive Time \%", "Avg. Elapsed Inclusive Time" and "Avg. Elapsed Exclusive Time" which are defined accordingly "The percentage of time spent executing a function and its child functions", "The percentage of time spent executing a function excluding its child functions", "The average time spent executing a function and its child functions" and "The average time spent executing a function excluding its child functions". Table \ref{tab:example} shows most used functions in each benchmark and the results gathered from instrumentation.
\begin{table*}[ht]
{\small
\hfill{}
\caption{The function reference computation using Visual Studio 2013 profiler instrumentation}
\makebox[18.5cm]{
\label{tab:example}
\centering
\begin{tabular}{|c|c|c|c|c|c|c|}
    \hline
Benchmarks&Function&Number &Elapsed Inclusive &Elapsed Exclusive&Avg. Elapsed &Avg. Elapsed\\
          & Name   & of Calls & Time\% &  Time\% & Inclusive Time (sec)& Exclusive Time (sec)\\
    \hline
\multirow{2}{*}{blackscholes}&  mainCRTStartup &  1 & 100 & 34.41 & 162.89 & 56.05  \\
						     & bs\_thread & 1 & 34 & 34 & 55.38 &  55.38 \\
	\hline
\multirow{2}{*}{specrand}&  printf &  1,002 & 99.9 & 99.9 & 0.15 & 0.15  \\
						 & mainCRTStartup & 1 & 99.98 & 0.05 & 152.25 & 0.08 \\
	\hline
\multirow{3}{*}{mm}&  printf &  8 & 51.93 & 51.93 & 0.06 & 0.06  \\
				   & pthread\_join & 2 & 21.39 & 21.39 & 0.1 & 0.1 \\
				   & pthread\_create & 2 & 14.76 & 14.76 & 0.07 &  0.07 \\
	\hline
\multirow{3}{*}{qs}&  printf &  5 & 43.91 & 43.91 & 0.1 & 0.1  \\
				   & pthread\_join & 2 & 24.74 & 24.74 & 0.14 & 0.14 \\
				   & pthread\_create & 2 & 17.12 & 17.12 & 0.1 &  0.1 \\
	\hline
\multirow{3}{*}{factorial}& printf &  1 & 78.09 & 78.09 & 0.23 & 0.23  \\
						 & CxxSetUnhandled & 1 & 9.86 & 9.86 & 0.03 & 0.03 \\
						 & ExceptionFilter& & & & &\\
	\hline
\multirow{2}{*}{circular\_buffer}&  pthread\_join &  2 & 47.17 & 47.17 & 0.12 & 0.12  \\
								 & pthread\_create & 2 & 35.41 & 35.41 & 0.09 & 0.09 \\
	\hline
\multirow{2}{*}{stack}&  pthread\_join &  2 & 55.09 & 55.09 & 0.19 & 0.19  \\
								 & pthread\_create & 2 & 32.52 & 32.52 & 0.11 &  0.11 \\
	\hline
\end{tabular}}
}
\hfill{}
\label{tb:profiler}
\end{table*}

\subsection{The Sensitivity Analysis of Instructions to Soft Errors}

We define the  application resilience to soft errors as its ability
to tolerate hardware faults if they occur, without leading
to an incorrect output. Incorrect outputs are also known as
Silent Data Corruptions (SDCs). To recover this group of failures, there is no dedicated method to indicate that the application has malfunctioned (unlike a
crash or a hang, where either an exception is raised or a timeout occurs). Since, we are primarily interested in evaluating the soft error immunity of applications, we only inject faults into the
program's data or instructions that are visible at the assembly code or higher levels, rather than into the micro-architectural structures. 

Accordingly, we classify the outcome of activated faults based on the program's behavior to following categories:
\begin{itemize}
\item Crash: if the program is terminated by the OS due to an exception.
\item Silent Data Corruption (SDC): if the output is incorrect due to lack of appropriate method to report the impact of the fault propagation.
\end{itemize}

\begin{figure}
    \centering
    \subfigure[]
    {
        \includegraphics[width=3.2in]{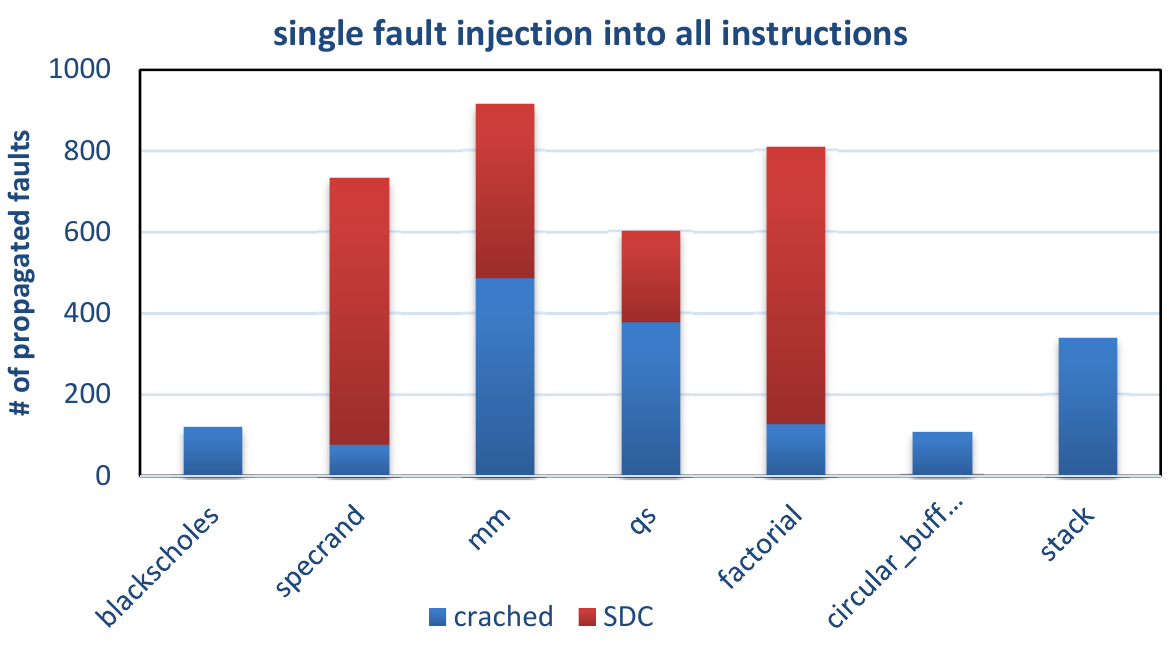}
       \label{fig:first_sub}
    }
    \caption{Aggregated single fault injection results with LLFI for all operation instructions}
    \label{fig:SEU}
\end{figure}

\begin{figure}
    \centering
    \subfigure[]
    {
        \includegraphics[width=3.2in]{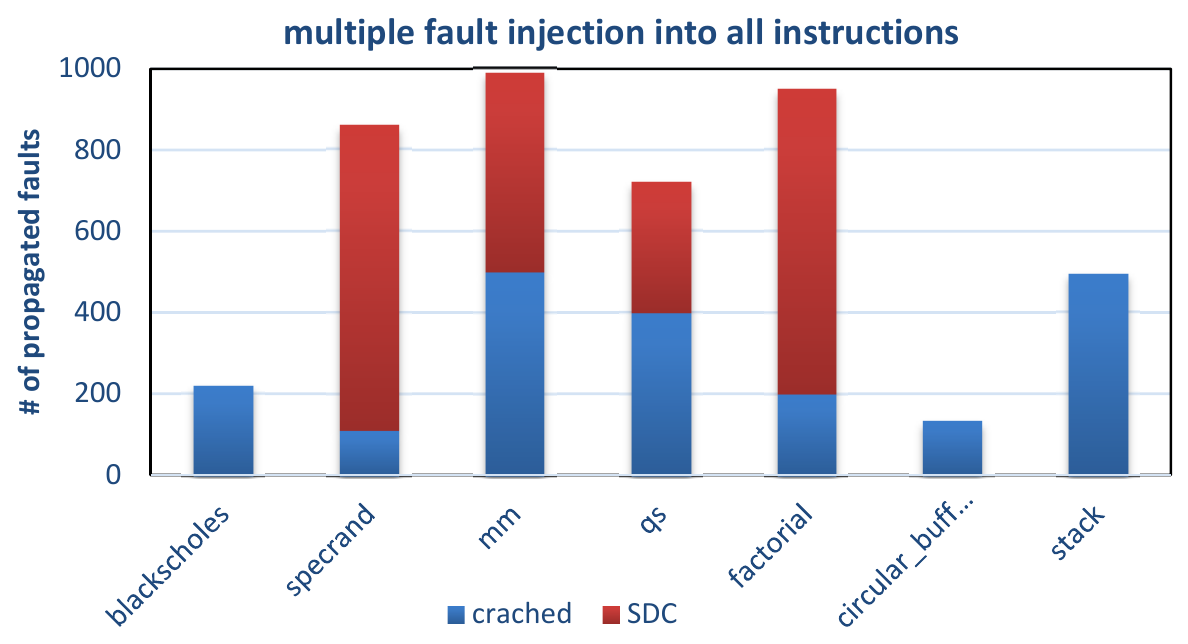}
       \label{fig:first_sub}
    }
    \caption{Aggregated multiple fault injection results with LLFI for all operation instructions} 
    \label{fig:MBU}
\end{figure}

Figure. \ref{fig:SEU} shows the impact of SEU on benchmarks suite which benefit from POSIX Pthreads standard \cite{butenhof1997programming} for creating and handling threads. The most SEUs in $specrand$ and $factorial$ benchmarks result in SDC in our system while other benchmarks often crash when a soft error occurs. Furthermore, Figure. \ref{fig:MBU} shows that the rate of crash or SDC significantly increase as the number of flipped bits increased. This results confirm our previous statement that MBUs are major reliability issue in current multi-core systems which demand a comprehensive solution for mitigating them.

To determine the most sensitive instructions to soft error, we target various type of instructions. We noticed that $pthread$ related code fragments show high vulnerability to soft errors. This means $pthread$ related code fragments are required to be mapped on soft-error-immune storage and operation components. For further information, please refer to Appendix A.

\section{The Proposed Execution Model}
To completely benefit from the intrinsic soft error resiliency characteristic of STT-RAM, we also need to thoroughly explore the other aspects of using STT-RAM instead of SRAM to make sure the requirements for performance specifications and energy constraints will be met. This section first describes the structure of STT-RAM and the steps associated to each read or write operation.  Then, our hardware/software model assumed in our system and the particular instruction protection strategies applies to our program is presented. 
\subsection{The Spin-Transfer Torque Magnetic Random Access Memory (STT-RAM)}
\begin{figure}
\centering
\includegraphics[width=3in, keepaspectratio]{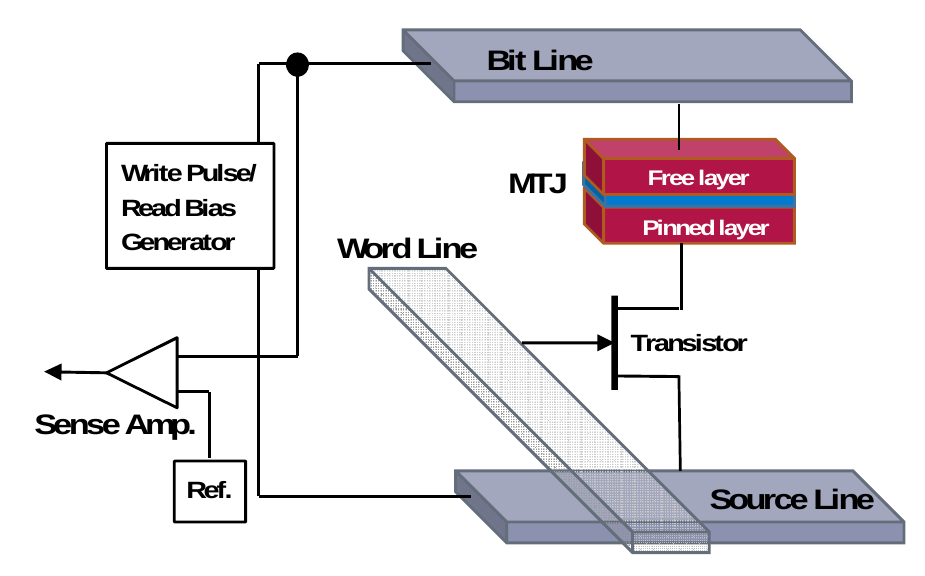}
\caption{An illustration of a 1T1J STT-RAM cell.}
\label{sttram}
\end{figure}

As illustrated in Figure. \ref{sttram}, the STT-RAM  uses  magnetic  elements  called  magnetic  tunneling junction (MTJ) to store data in which a thin insulating oxide later, e.g. MgO, is sandwiched by two ferromagnetic layers. Moreover, the upper ferromagnetic layer usually aliased as free layer, its polarity of magnetic field can be flipped over during a write event; while, the lower ferromagnetic layer usually called as pinned layer is designed to have its magnetization pinned. Thus, MTJ has low (high) resistance distribution if the magnetization of the free layer and the pinned layer are aligned (anti-aligned). Accordingly, low (high) resistance distribution is stored in MTJ, instead of traditional electronic charge or current flow. 

\begin{figure}
\centering
\includegraphics[width=3in, keepaspectratio]{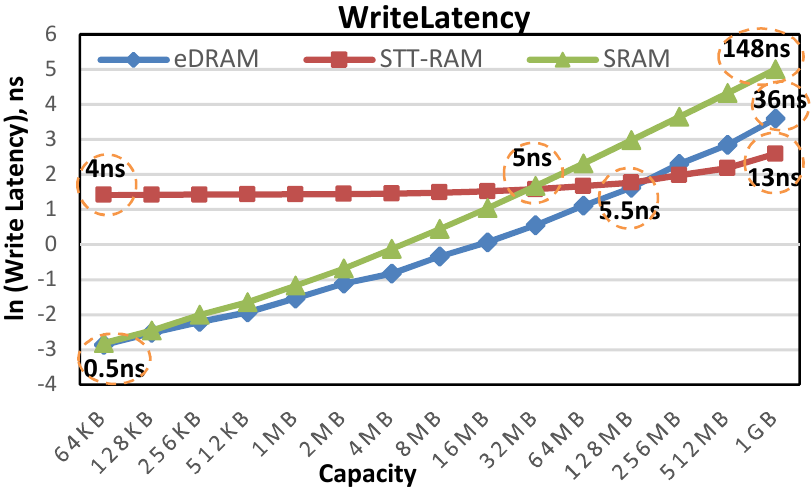}
\caption{The write latency of STT-RAM compared to other memory technologies.}
\label{sttram_write}
\end{figure}

For a read operation, a small current is required to be driven from bit-line to source line. Unlike read operation, a successful write operation requires a current flow drive either from bit-line to source-line or vice versa, depending on the differential voltage between these two lines. Although STT-RAM does not suffer from write endurance, however the advent of long write latency and high energy consumption exacerbate the performance of STT-RAM. Figure. \ref{sttram_write} shows the write latency comparison for various cache module configurations among three well-known memory technologies including eDRAM, STT-RAM, and SRAM obtained from NVSim \cite{nvsim} for $45nm$ technology node. This comparison shows that the STT-RAM is not a good candidate for small size memory module due to its long write latency compared to other technologies. This means that the design of proposed architecture should carefully leverage the potential of STT-RAM for small size storage elements like register arrays in the processor's pipeline.  

\subsection{Soft Error Resilient Architecture Configuration }
We want to make both highly referenced storage elements and operations in the proposed architecture immune to soft error strikes. The proposed hardware and software interface is built on the previous works that enable programmers to map nonvolatile data on nonvolatile main memory \cite{coburn2011nv} \cite{volos2011mnemosyne}. These techniques consist of language, compiler, and runtime system support to manage nonvolatile data. We extended the previous frameworks to allow programmers to use special keywords and library calls to handle data that require soft-error-immune storage component. In particular, we used the sensitivity analysis of instructions to soft errors and the rate of instruction referencing to determine which typeset of instructions are required to be mapped on soft-error-resilient components. 
   
\begin{figure*}
\centering
\includegraphics[width=7in, keepaspectratio]{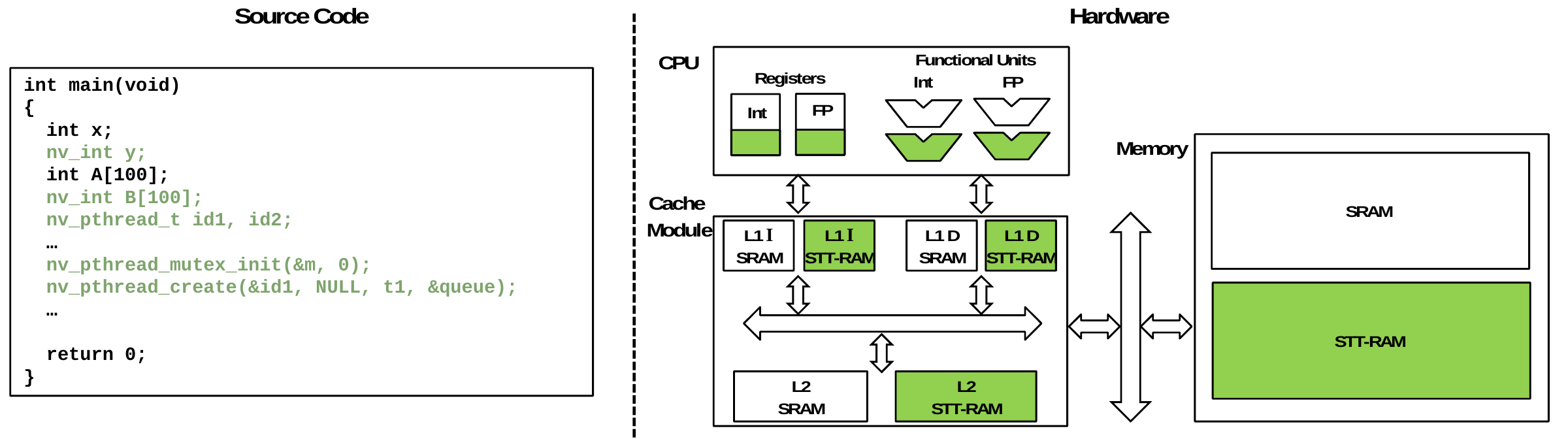}
\caption{HW/SW model assumed in our system. The green areas are made of soft-error-resilient memory module.}
\label{HW_model}
\end{figure*}

As illustrated in Figure. \ref{HW_model}, the proposed storage is offered in the form of both reliable and unreliable storage components. The reliable storage components are made of STT-RAM offering high resiliency to soft errors while the traditional SRAM cells are used for creating unreliable storage elements. To be specific, the reliable and unreliable registers are distinguished based on the register number. The reliable data stored in memory is distinguished from unreliable data based on regions of physical memory address. This can be done by the proposed approach in \cite{nvm_duet2014} where the reliable data can be linked to a reserved virtual address space (reliable space). When this reliable address is accessed, the data is stored in reliable portions of the data cache. Note that the hybrid SRAM and STT-RAM cache configuration techniques \cite{sun2011multi} \cite{wang2014coherent} have been proposed in the past and can be utilized to allocate data across the hybrid cache hierarchy.

For reliable operations, a specific keywords and library calls such as "$nv\_pthread\_t$" (a nonvolatile version of $pthread\_t$) are available to programmers for reliable integer ALU operations as well as reliable floating point operations. The reliable instructions can use special functional units which are made of STT-RAM offering soft error resiliency. Note that even if a field which should be mapped into soft-error-immune field ends up stored in an unreliable memory, it will still be loaded into reliable registers and be subject to reliable operations.
 
\section{Experimental Results}
In order to evaluate the proposed technique, LLVM-based fault injection tool called $LLFI$ \cite{wei2014quantifying} has been used to inject transient faults into the multithreaded programs in a multi-core system (Intel(R) Core(TM) i7-4770 CPU @ 3.40GHz, RAM=6GB, OS=Linux Ubuntu 14.04.3 LTS). LLFI works at the LLVM compiler's IR level, and allows fault-injections to be performed at specific program points, and into particular instructions. LLFI
supports various fault injection customizations, and enables tracing the propagation of the fault among instructions in the program.

The steps required to inject faults using LLFI are as following:
\begin{itemize}
\item In Step 1, LLFI takes the program IR as input, and applies custom fault injection instruction and operand(s) selector to determine which instructions/operands are fault
injection candidates.
\item In Step 2, LLFI instruments the fault
injection instructions/operands with calls to fault injection
functions. The fault injection functions are designed to perturb
the specific instruction operand according to the specified
fault type at runtime (e.g. flip one bit of the operand for
bit-flip faults).
\item In Step 3, the compiled program is executed
at runtime, and LLFI randomly selects one runtime instance
of the instrumented instructions to trigger the fault injection
function and inject into the selected instruction operand value.
\end{itemize}
  
Because hardware faults occur randomly at runtime, LLFI
picks a random instruction from the set of all dynamically
executed instructions at runtime to inject into. 

The benchmarks suite we used in our experimental results are as following:
\begin{itemize}
\item blackscholes: The blackscholes application is an Intel RMS benchmark. It calculates the prices for a portfolio of European options analytically with the Black-Scholes partial differential equation.
(PDE).
\item specrand: The benchmark simply generates a sequence of pseudorandom numbers starting with a known seed. 
\item Matrix Multiplication (mm): It is a simple matrix multiplication program in which the main thread makes slave threads responsible to compute each elements of the product separately and concurrently.
\item Quick Sort (qs): It is a simple quick sort program in which main thread first partitions the 100-elements array of integers into two parts, by performing one round of the quick sort algorithm, then assigns each sub-arrays to a slave thread in order to sort each part separately and simultaneously.
\item factorial: It computes the product of all positive integers less than or equal to $n$.
\item circular\_buffer: It simulates a buffer using shared variables to synchronize receive and send operations. 
\item stack: It is a program simulating  a data-stack structure. 
\end{itemize}

\begin{figure}
\centering
\includegraphics[width=3in, keepaspectratio]{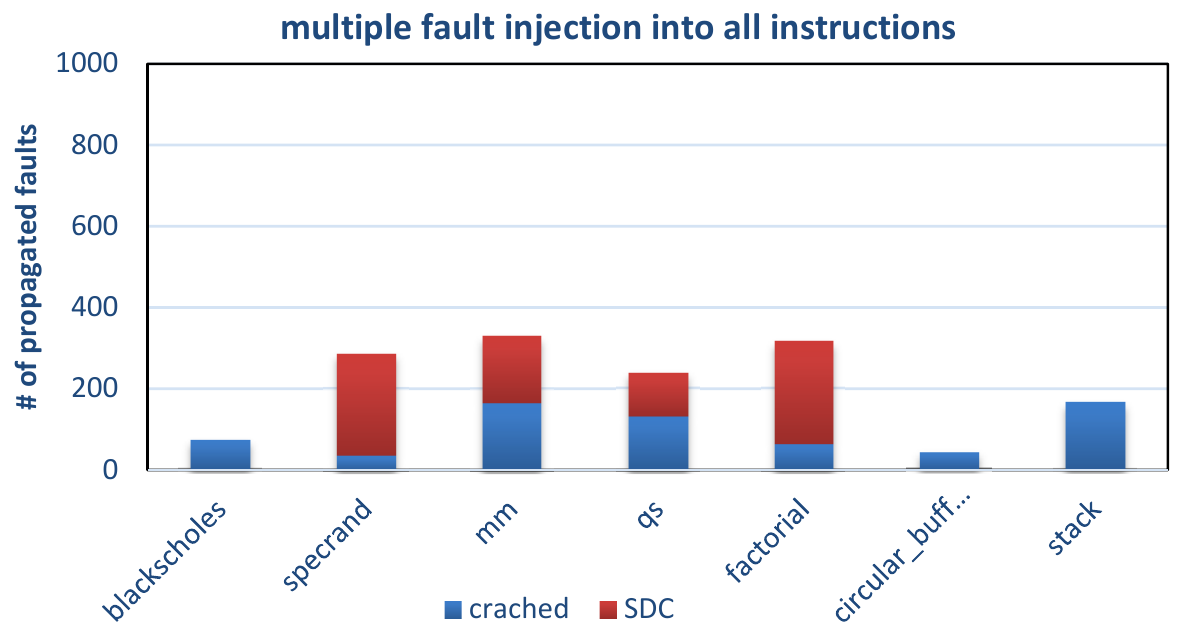}
\caption{Aggregated results for multiple fault injection into all operation instructions of the proposed HW model.}
\label{nv_MBU_all}
\end{figure}

\begin{figure}
\centering
\includegraphics[width=3in, keepaspectratio]{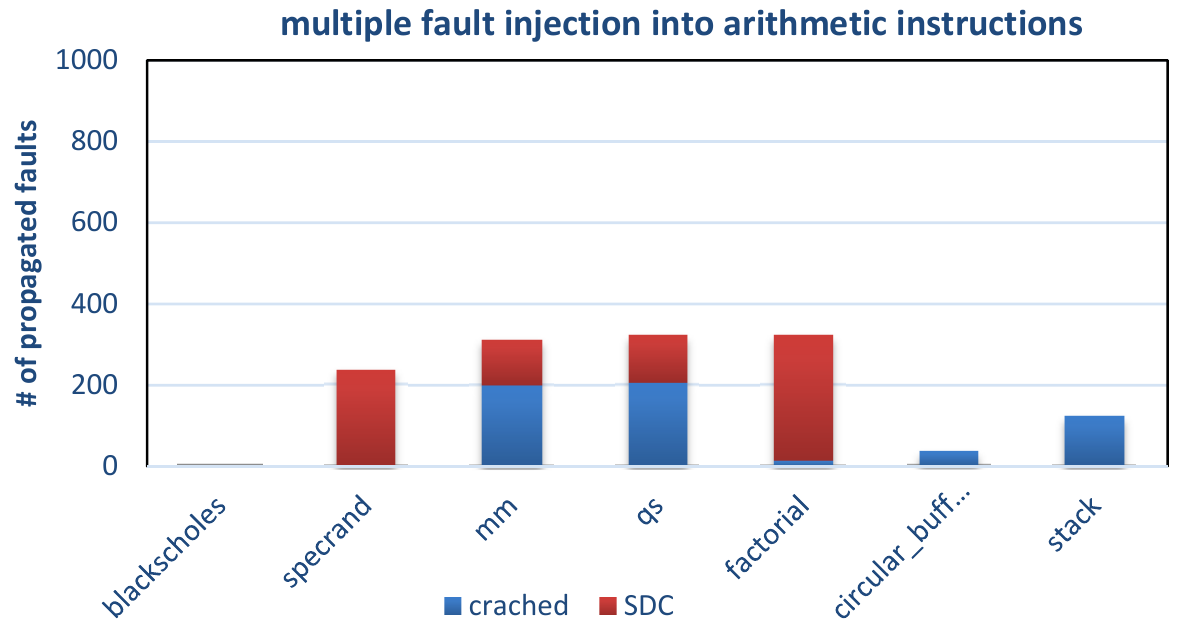}
\caption{Aggregated results for multiple fault injection into arithmetic operation instructions of the proposed HW model.}
\label{nv_MBU_arithmetic}
\end{figure}

\begin{figure}
\centering
\includegraphics[width=3in, keepaspectratio]{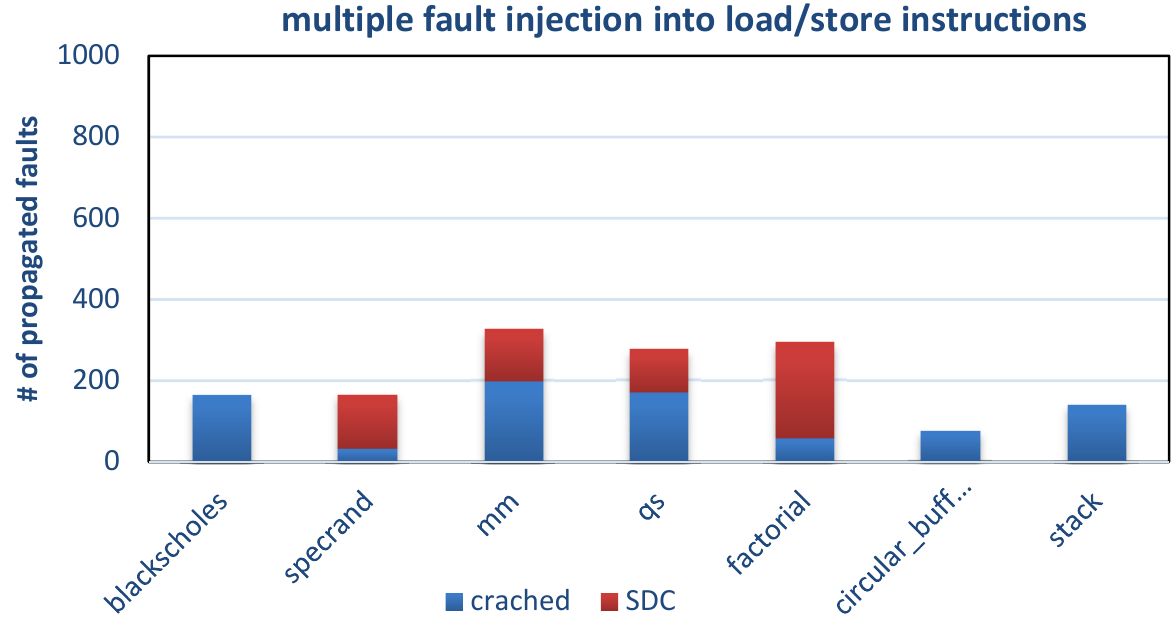}
\caption{Aggregated results for multiple fault injection into load/store operation instructions of the proposed HW model.}
\label{nv_MBU_load_store}
\end{figure}

Since we are interested in the study of the impact of MBUs in the existing CMP architectures, we injected 1000 multiple faults into the assumed hardware model in our system. The results show that the proposed architecture design is 30\% on average more resilient to soft errors as shown in Figure. \ref{nv_MBU_all}, \ref{nv_MBU_arithmetic} and \ref{nv_MBU_load_store}. The obtained results show the efficiency of the proposed method to handle soft errors in the entire CMP architecture. For most of the benchmarks, the replacement of a portion of memory modules reduces the period during which the data in the storage component are exposed to particle strikes. In the cache hierarchy, the use of STT-RAM to maintain frequently read cachelines in the high-dense low-level caches significantly increase the reliability level of the system. The main reason for the high vulnerability of traditional low-level caches to soft error is the high potential of residing a data block in the last level cache for millions of cycles between two consecutive accesses \cite{ref:Loh_2011}. 

In the pipeline stage, the replacement of a portion of unreliable registers and functional units with reliable elements eliminate a high portion of faults in operation components. However, the soft error in logic components still can be propagated in our proposed approach. Nonetheless, this portion of faults are relatively small compared to other faults which exclusively target storage and operation components.
  
\section{Conclusions}
In this paper, we focused on leveraging the intrinsic soft-error-immunity characteristic of STT-RAM as an alternative for SRAM-based storage and operation components. We showed that a specific portion of working set in the multithreaded programs are ideal candidate to be stored and executed in STT-RAM based components. Doing so, the proposed CMP architecutre can achieve 30\% on average more resilient to soft errors.

\section{Future Work}
We still look for a methodology to determine those instructions which experience high rate of references with the lowest memory modification. To attain this goal, we started to look into $intel~vtune$ \cite{moseley2007identifying} which provides a rich set of performance insight into CPU performance, threading performance \& scalability, bandwidth, caching and much more. We expect to better determine the candidates for using STT-RAM elements through obtaining the rate of memory modification by executing each instruction. In addition, another section in experimental results needs to be added which shows the amount of energy consumption and performance benefit achieved using the proposed method compared to traditional approach.

%
\bibliographystyle{abbrv}
\bibliography{sigproc}  
%
\section{Appendix A}
\begin{figure}
    \centering
		\subfigure[]
    {
        \includegraphics[width=3.2in]{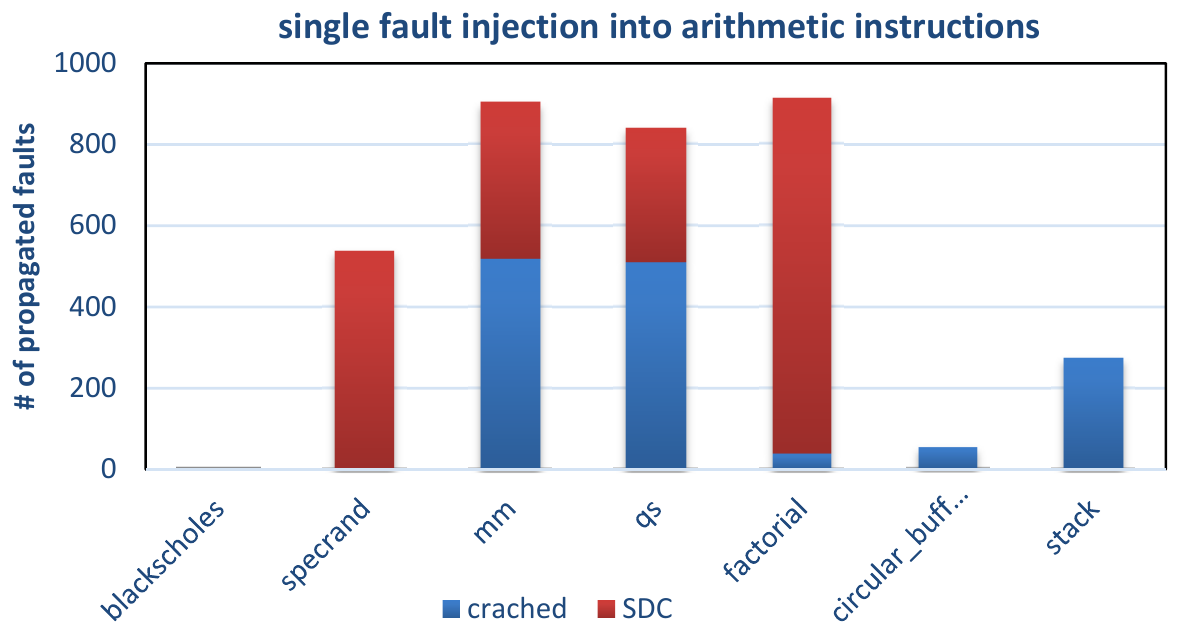}
        \label{fig:second_sub}
    }
    \subfigure[]
    {
        \includegraphics[width=3.2in]{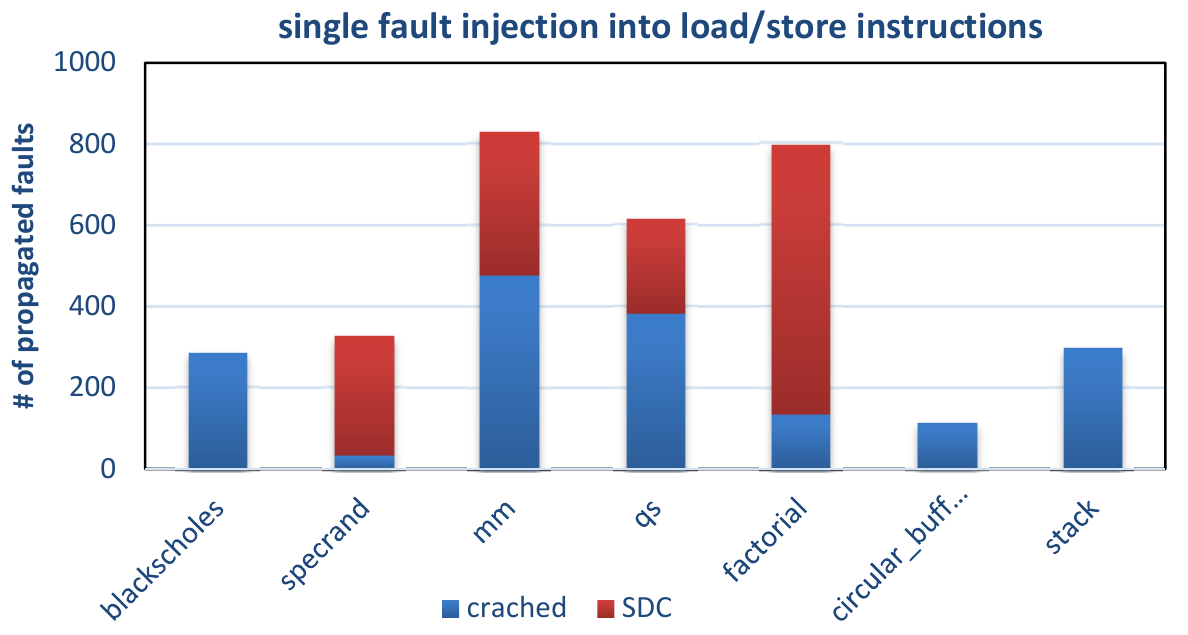}
        \label{fig:third_sub}
    }
    \caption{Aggregated single fault injection results with LLFI for all operation instructions (a)  Arithmetic operation instructions, (b)  Load/store instructions}
    \label{fig:SEU2}
\end{figure}

\begin{figure}
    \centering
		\subfigure[]
    {
        \includegraphics[width=3.2in]{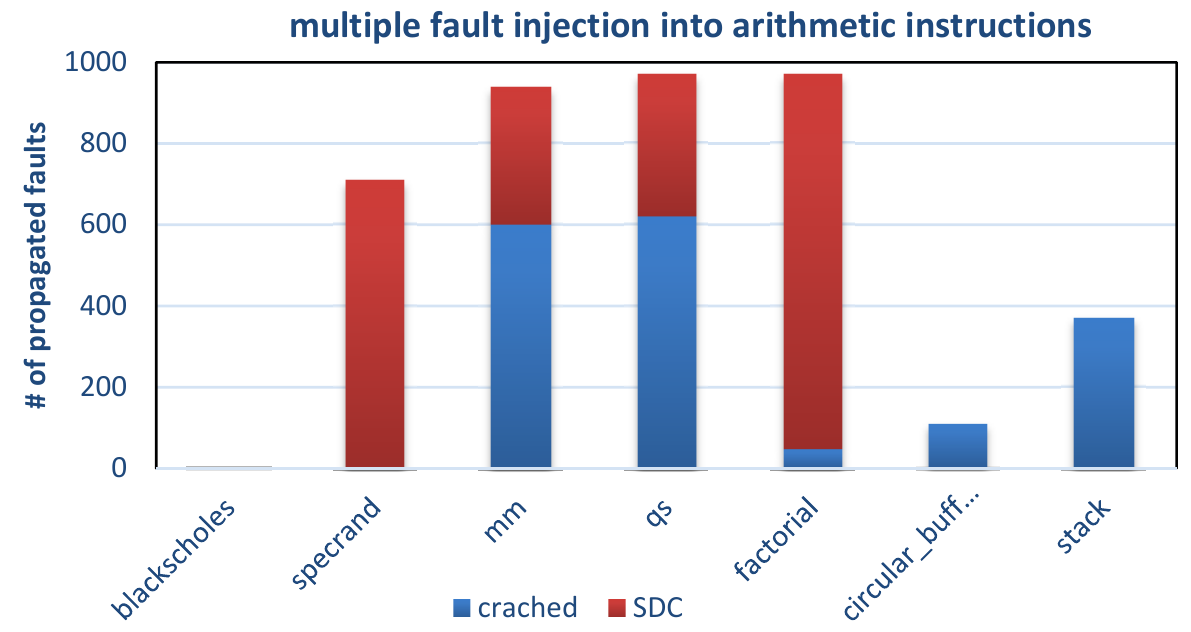}
        \label{fig:second_sub}
    }
    \subfigure[]
    {
        \includegraphics[width=3.2in]{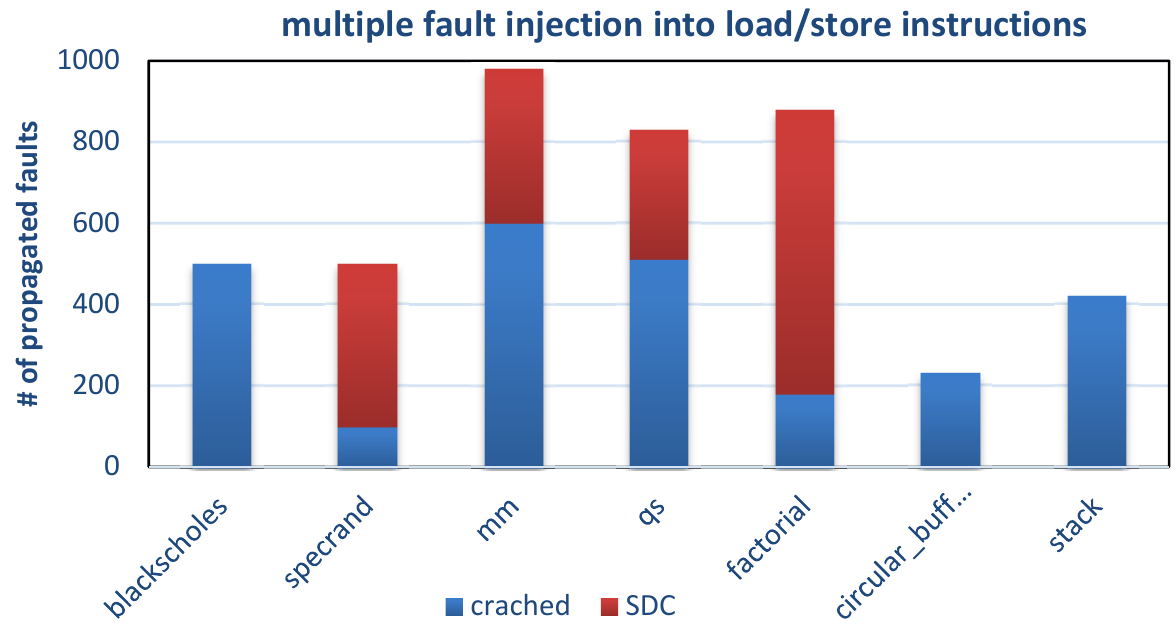}
        \label{fig:third_sub}
    }
    \caption{Aggregated multiple fault injection results with LLFI for (a)  Arithmetic operation instructions, (b)  Load/store instructions}
    \label{fig:MBU2}
\end{figure}

A detail study of the impact of soft errors on multithreaded programs are shown in Figure. \ref{fig:SEU2} and \ref{fig:MBU2}. The soft errors contribute the most SDCs in $specrand$ and $factorial$ benchmarks in both arithmetic and load/store instruction typesets while other benchmarks only crash depend on the sensitivity of their instructions to soft errors. For example, $mm$, $qs$ and $stack$ benchmarks show high sensitivity to soft errors while the $circular\_buffer$ workload does not readily get influenced by soft errors.  

\end{document}